\documentclass[12pt ,a4paper]{article}
\usepackage{titlesec}

\titleformat{\chapter}{\normalfont\huge}{\thechapter.}{20pt}{\huge\textbf}

\usepackage[english]{babel} 
\usepackage[T1]{fontenc}
\usepackage{makeidx}
\usepackage{appendix}
\usepackage{amsfonts}
\usepackage{amsmath, relsize}
\usepackage{amsmath}
\usepackage{slashed}
\usepackage{amssymb}
\usepackage{graphicx}
\usepackage{mathrsfs}
\usepackage{amsthm}
\usepackage{amssymb}
\usepackage{verbatim}
\usepackage{booktabs}
\usepackage{pxfonts}
\usepackage{cite} 
\usepackage{setspace} 
\usepackage{hyperref}
\hypersetup{
     colorlinks   = true,
     citecolor    = black,
     linkcolor   = black
}

\usepackage{caption}
\usepackage{subcaption}

\usepackage{float} 

\author{Chiara Gastaldi}
\date{\today}

\begin{document}

\begin{center}
\begin{large}
\textbf{Optimal stimulation protocol in a bistable synaptic consolidation model}\\
\end{large}
\vspace{0.4cm}
Chiara Gastaldi, Samuel P. Muscinelli and   Wulfram Gerstner\\
\vspace{0.4cm}
\small{\'Ecole Polytechnique
F\'ed\'erale de Lausanne, CH-1015 Lausanne, Switzerland
}
\vspace{0.1cm}
\end{center}

\vskip 0.2cm
\begin{abstract}
Consolidation of synaptic changes in response to neural activity is thought to be fundamental for memory maintenance over a timescale of hours.
In experiments, synaptic consolidation can be induced by repeatedly stimulating presynaptic neurons.
However, the effectiveness of such protocols depends crucially on the repetition frequency of the stimulations and the mechanisms that cause this complex dependence are unknown.
Here we propose a simple mathematical model that allows us to systematically study the interaction between the stimulation protocol and synaptic consolidation.
We show the existence of optimal stimulation protocols for our model and, similarly to LTP experiments, the repetition frequency of the stimulation plays a crucial role in achieving consolidation.
Our results show that the complex dependence of LTP on the stimulation frequency emerges naturally from a model which satisfies only minimal bistability requirements.

\end{abstract}

\section{Introduction}
\label{intro}

Synaptic plasticity, i.e. the modification of the synaptic efficacies due to neural activity,  is considered the neural correlate of learning \cite{hebb1949,hayashi2015,holtmaat2016,caroni2012,nabavi2014engineering}.
It involves several interacting biochemical mechanisms, which act on multiple time scales.
The induction protocols for short-term plasticity (STP, on the order of hundreds of milliseconds) \cite{markram1998differential} and for the early phase of long-term potentiation  or depression (LTP or LTD, on the order of minutes to hours) are well established 
\cite{bienenstock1982,pfister2006,fremaux2015, clopath2010, gjorgjieva2011}.
On the other hand, various  experiments have shown that, on the timescale of hours, the evolution of synaptic efficacies depends in a complex way on the stimulation protocol \cite{dudai2000, nader2000, redondo2011, frey1997}. 
This phenomenon is called \emph{synaptic} consolidation, to be distinguished from \emph{memory} consolidation, which is believed to take place through the interaction between hippocampus and cortex and which occurs on an even longer timescale  \cite{hasselmo1999, brandon2011, roelfsema2006}.
Such a richness of plasticity mechanisms across multiple time scales has been hypothesized to play an important role in increasing the capacity of synapses \cite{fusi2005, benna2016}.

Synaptic consolidation is often studied in hippocampal or cortical slices, in which it is induced by extra-cellular stimulation of afferent fibers \cite{frey1997,sajikumar2004late,sajikumar2004resetting}, which usually consists of a  series of short current pulses of variable amplitude and frequency. 
Experimental protocols are often organized in multiple repetitions of such stimulations, with  variable repetition frequency and duration of each episode. 
The dependence of the consolidation dynamics on the parameters of the experimental protocol is complex and has remained elusive. 
Both the current pulse frequency and the inter-episode delay play an important role in determining whether a synapse gets potentiated or not after the stimulation, and there is evidence that supports the existence of optimal parameter set to achieve potentiation \cite{larson2015theta}.   
Existing models succeeded in reproducing  experimental results on early and late LTP
\cite{barrett2009, clopath2008, ziegler2015}, providing an insightful mathematical description of the interaction of different mechanisms in a synapse.
However, the complexity of those models prevented their complete dynamical characterization, effectively not improving our understanding of the relationship between input and synaptic consolidation.
Here we address the following question: why is the temporal structure of stimulation, i.e. the timing of repetitions, so important for synaptic consolidation?

We introduce a simplified model of synaptic consolidation,  whose dynamics are rich enough to show bistability and transitions between  depotentiated and potentiated states of single synaptic contact points, as it is supported by  experimental results \cite{petersen1998all, O'Connor2005}.
We find that, despite the simplicity of our model, potentiation of a synapse depends in a complex way on the temporal profile of the stimulation protocol.
Our results suggest that not just the number of pulses, but also the precise timing within and across repetitions of a stimulation sequence, can be important, in agreement with anecdotal evidence that changes in protocols can have unexpected consequences.

\section{Methods}
In this section, we introduce and motivate the synaptic consolidation model that we analyze in the Results section.
Since describing the details of molecular interactions inside a synapse as a  system of differential equations \cite{lisman2001model,bhalla1999emergent}  would be far too complicated for our purpose, we aim to capture the essential dynamics responsible for consolidation experiments with an effective low-dimensional dynamical system.
In this view, auxiliary variables are mathematical abstractions that might represent the average effect of a network of biochemical cascades, e.g. during a transition from one metastable configuration to another.

A one-dimensional dynamical system is not expressive enough to capture experimental data.
Indeed, in a one-dimension differential equation, it would be sufficient to know the instantaneous state of a single variable of the synapse (such as the weight) to predict its evolution, while this is not the case in experiments.
As a natural step forward, we consider a general autonomous two-dimensional model
\begin{equation}
    \begin{split}
        \frac{dw}{dt}&=f(w,z)\\
        \frac{dz}{dt}&=g(w,z) \quad ,
    \end{split}
    \label{system}
\end{equation}
where $w$ represents the measured efficacy of a synaptic contact point (e.g. the amplitude of the EPSP caused by  pre-synaptic spike arrival), while $z$ is an abstract auxiliary variable.
For simplicity, both variables will be considered unit-less.
One way to tackle the very general system in Eq. (\ref{system}) is to perform a Taylor expansion around $(w,z)=(0,0)$. 
\begin{eqnarray}
&&\frac{dw}{dt}=A(z)+B(z)\cdot w+C(z)\cdot w^2 + D(z)\cdot w^3 + \dots \\
&&\frac{dz}{dt}=A'(w)+B'(w)\cdot z + C'(w)\cdot z^2 + D'(w)\cdot z^3 \dots \quad .
\end{eqnarray}
We consider the situation in which we have linear coupling between the two variables, i.e. $A(z)=A_0+A_1\cdot z$, $B(z)=B$, $C(z)=C$ and $D(z)=D$.
Analogously, in the second equation we set $A'(w) = A_0' + A_1'\cdot w$, $B'(w)=B'$, $C'(w)=C'$ and $D'(w)=D'$.
The choice to expand up to the third order is due to the fact that, to implement the bistable dynamics \cite{petersen1998all, O'Connor2005} of single contact points in a dynamical system approach, we have two requirements: 1) the systems should have at least two stable fixed points
and 2) the dynamics should be bounded.
These conditions cannot be met by degree 1 or degree 2 polynomials since they can have at most one \textit{stable} fixed point.
Therefore bistability requires a degree 3 polynomial in at least one equation.
To be more general, we will consider a system in which both the polynomials can be of degree 3.

In the following, we restrict to the case in which we have at least two stable fixed points in the two-dimensional, third order, bounded dynamical system. This can be obtained with a negative coefficient of the third power in both equations.
Assuming that the degree 3 polynomial has three real roots, we can rewrite our system in the more intuitive form
\begin{equation}
\begin{split}
\tau_w \frac{dw}{dt} &= -K_1(w-w_1)(w-w_2)(w-w_3) + C_1 z \\
\tau_z \frac{dz}{dt} &= -K_2(z-z_1)(z-z_2)(z-z_3) + C_2 w \quad ,
\end{split}
\end{equation}
where $C_1$ and $C_2$ are coupling constants and the roots correspond to the fixed points of the equations in the uncoupled case ( $C_1=C_2=0$).
$\tau_w$ and $\tau_z$ can be interpreted as time constants since they do not influence the location of the fixed points but only the speed of the dynamics.
$K_1$ and $K_2$ are two positive constants that scale the whole polynomial, while $C_1$ and $C_2$ are positive constants that control the amount of coupling between the two variables.
If we exclude the coupling terms, each equation corresponds to an over-damped particle moving in a double-well potential \cite{strogatz2014nonlinear}.

In order to further simplify our study, we assume that in both equations one of the roots is zero and that the other two roots are one positive and one negative, equally distant from zero.
Notice that we are assuming that a synaptic weight can take both positive and negative values, which is not biologically plausible.
However, this choice simplifies the calculations without loss of conceptual insights, since it is always possible to go back to a system with  positive weights by applying a coordinate translation.
Following \cite{zenke2015}, we add a plasticity induction term to the first equation that describes the driving by an LTP induction protocol.
The equations now read
\begin{equation}
\begin{split}
\tau_w \frac{dw}{dt} &= -K_1(w-w_0)(w+w_0)w + C_1 z + I \\
\tau_z \frac{dz}{dt} &= -K_2(z-z_0)(z+z_0)z + C_2 w \quad .
\end{split}
\label{eq:potential}
\end{equation}
In the absence of coupling, the double well potential, related to Eq. \ref{eq:potential} has minima in $w =\pm w_0$, $z =\pm z_0$ and a local maximum in $w=0$ ($z=0$).

The system has  nine free parameters.
However, since the stable fixed points of the system are easier to access experimentally (the value of $w$ at the stable fixed point should be related to the synaptic weight measured experimentally), it is useful to rewrite the system as
\begin{equation}\label{eq:dynamics}
\begin{split}
\tau_w \frac{dw}{dt} &= -K_w(w-w_0)(w+w_0)w 
+ C_w \left( z - \frac{z_0}{w_0} w \right) + I \\
\tau_z \frac{dz}{dt} &= -K_z(z-z_0)(z+z_0)z 
+ C_z \left( w -\frac{w_0}{z_0}z\right) \quad .
\end{split}
\end{equation}
In this way, $(w,z) = \pm(w_0,z_0)$ are the stable fixed points of the two-dimensional system.
For $K_w\neq0$ and $K_z\neq0$, we could divide Eq. \ref{eq:dynamics} by   $K_w$ and $K_z$ to further reduce the numbers of parameters. However, we will stick to a notation with explicit $K_w$ and $K_z$ but, without loss of generality, we will choose $K_{w,z}\in \{0,1\}$.
Note that the choice $K_z=0$ implies that the dynamics of the auxiliary variable $z$ are linear.

Since the system is two-dimensional, it can be studied using phase-plane analysis. 
The fixed points of the system are graphically represented by the intersections of the nullclines (i.e. the curves defined by either $\frac{dw}{dt}=0$ or $\frac{dz}{dt}=0$), which in our system are:
\begin{equation}
    \begin{split}
        w-\text{nullcline:} \,\,\,\, & 
        z=\frac{z_0}{w_0}w + \frac{K_w}{C_w}(w-w_0)(w+w_0)w - \frac{I}{C_w} \\
        z-\text{nullcline:} \,\,\,\, & 
        w=\frac{w_0}{z_0}z + \frac{K_z}{C_z}(z-z_0)(z+z_0)z \quad .
    \end{split}
\label{eq:nullclines}
\end{equation}
The maximum number of fixed points for the system of equations Eq. (\ref{eq:dynamics}) can be easily computed.
To do so, consider a more general form of two nullclines:
\begin{equation}
    \begin{split}
        w-\text{nullcline:} \,\,\,\, & z=P_n(w)\\
        z-\text{nullcline:} \,\,\,\, & w=Q_m(z) \quad ,
    \end{split}
\label{poly}
\end{equation}
where $P_n(z)$ is a polynomial of degree $n$ in $w$ and, analogously , $Q_m(w)$ is a polynomial of degree $m$ in $z$. To find the fixed points of the system we need to solve (\ref{poly}):
\begin{equation}
    w=Q_m(P_n(w)) \quad .
    \label{w_null}
\end{equation} 
Eq. (\ref{w_null}) is a polynomial equation of degree $n\cdot m$  in $w$ and therefore it allows a number of real solutions $s$, $0\leq s \leq n \cdot m$.
Applying this formula to our case, we find that we can have a maximum of nine fixed points.

We will start by considering the symmetric case (section \ref{subsect:symm_case}) in which the two equations have the same parameters, reducing the number of free parameters to five. 
Moreover, since $w_0 $ is related to measurable quantities, and since we make the choice $z_0=w_0=1$, the actual number of free parameters is four.
In the first subsection, we show the effect of changing the coupling coefficients. 
Then, we briefly comment on the effect of changing the time constants and on the effect of the induction current.
We will move to the analysis of the asymmetric cases in subsection \ref{subsect:asymm_case}.

\subsection{Symmetric changes of coupling coefficients reveal two bifurcations}
\label{subsect:symm_case}

\begin{figure}
\centering
\includegraphics[width=1.0\textwidth]{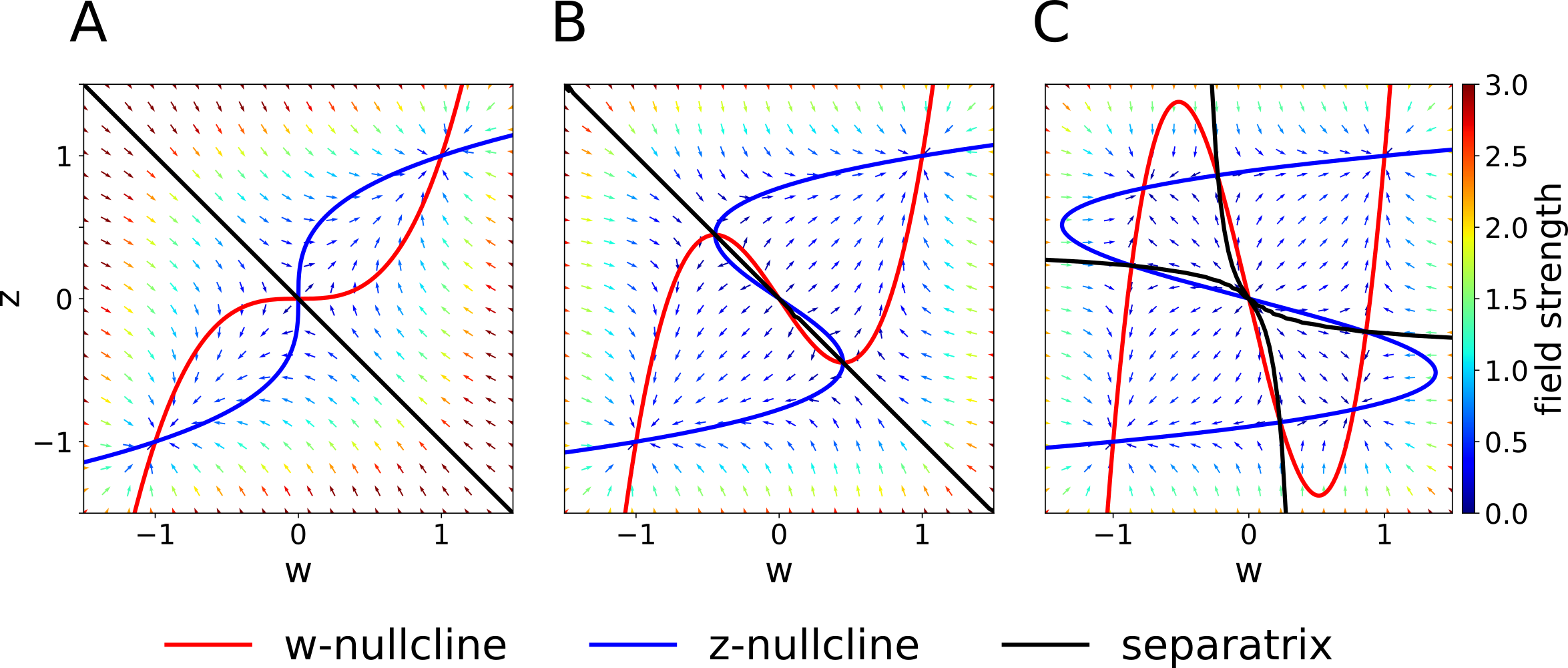}
\caption{ \textbf{Phase-plane diagram and basins of attractions for the symmetric case with  equal coupling constants, $C_w=C_z=C$}. The induction current is null, $I=0$.  \textbf{A:} $C=1$, phase-plane with field arrows. The  color of the arrows is proportional to the field strength. $w-$ and $z-$ nullclines are indicated in red and blue respectively. The line that separates the two basins of attraction is indicated in black. \textbf{B:} Same as A, but $C=0.4$.  Compared to A, we notice the creation of two  saddle points.  \textbf{C:} Same as A, but $C=0.2$. The maximum number of fixed points is achieved. In this case we have four basins of attraction.}
\label{fig:pp1}
\end{figure}

In this section, we  study the case of symmetric coupling $C_w=C_z=C$ and analyze how  a change of coupling strength influences the dynamics of the system.
As a side note, when the coupling is symmetric we can define a pseudopotential \cite{Cohen} 
\begin{equation}
V(w,z) = \frac{K_w}{4} w^4 + \frac{K_z}{4} z^4 
-\frac{1}{2w_0}\left(K_w w_0^3 -z_0C\right)w^2
-\frac{1}{2z_0}\left(K_z z_0^3 -w_0C\right)z^2
-Cwz
\end{equation}
in which the dynamical variables move according to $\frac{dw}{dt}=-\frac{\partial V}{\partial w}$ and  $\frac{dz}{dt}=-\frac{\partial V}{\partial z}$.

We fix $\tau_w=\tau_z$, $K_w=K_z=1$, $I=0$, $w_0=z_0=1$ and change $C$ in Eq. (\ref{eq:dynamics}).
In the case $C=1$, the system is in a rather simple regime: there are two stable fixed points in  $(w,z)=(-1,-1)$ and $(w,z)=(1,1)$ and a saddle fixed point in the origin (Fig. \ref{fig:pp1}).
 The basins of attraction of the stable fixed points are divided by the $z=-w$ diagonal.

\begin{figure}
\begin{center}
\includegraphics[width=0.70\textwidth]{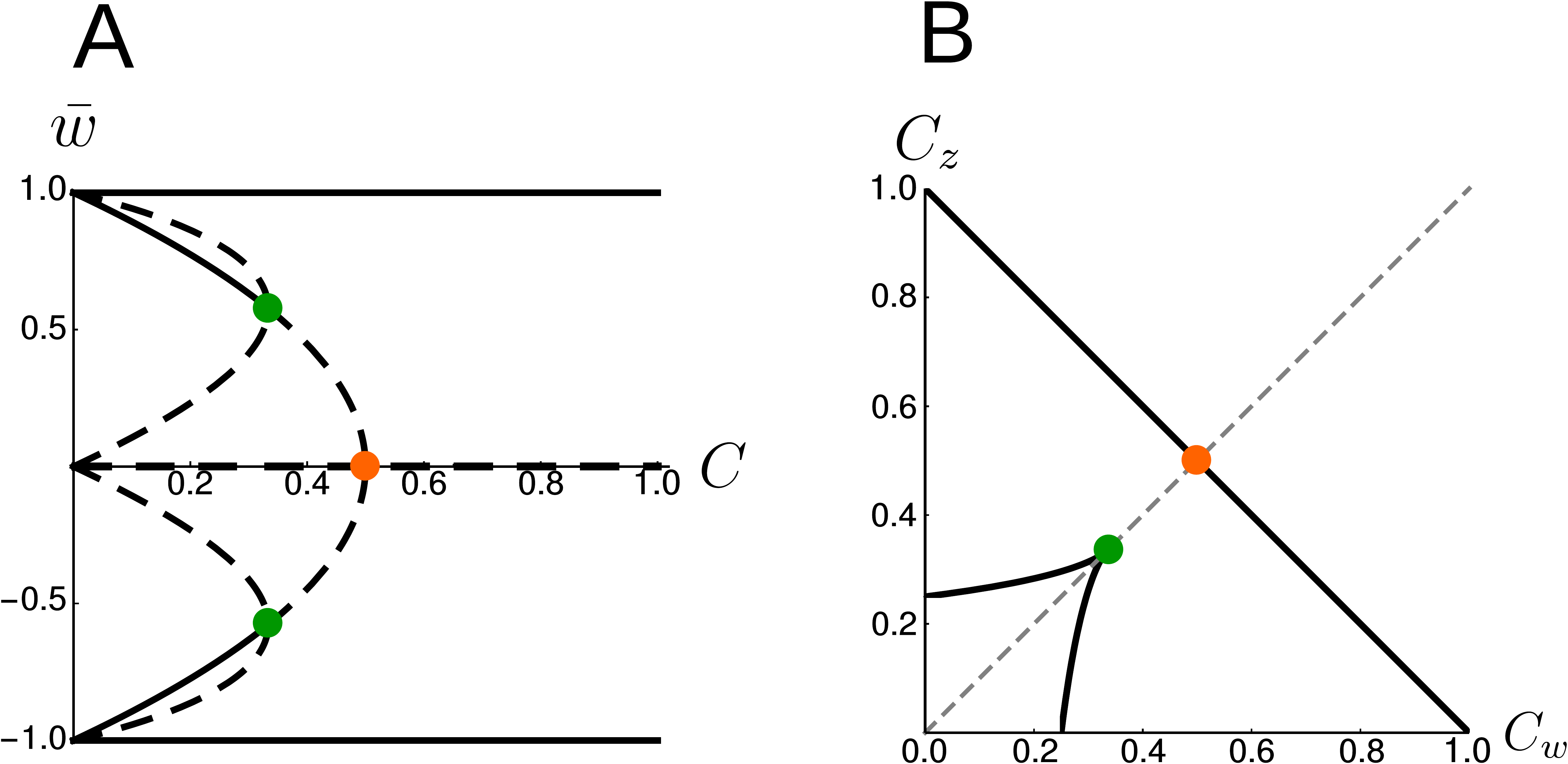}
\end{center}
\caption{\textbf{Bifurcations diagrams.} \textbf{A:} Fixed point positions in the symmetric case. Dashed lines indicate unstable fixed points while continuous lines indicate stable fixed points. 
Orange and green disks indicate bifurcation points.
\textbf{B:} Bifurcation points in the $C_w$-$C_z$ plane (black) for the general (asymmetric) case. The dashed gray line corresponds to $C_w=C_z$. The orange and green dots indicate the corresponding bifurcations in A. Note that, in B, the bifurcation at $C_w=C_z=\frac{1}{3}$ (green dot) is a degenerate point.}
\label{fig:bifurcations}
\end{figure}

If we decrease the coupling $C$, we encounter two bifurcations.
A first pitchfork bifurcation takes place at $C=1/2$, i.e. when the two nullclines are tangent to each other in the saddle fixed point. If we further decrease the value of the coupling coefficient, we can observe the creation of two additional saddle points (Fig. \ref{fig:pp1}.B). 
 The stability properties of the other fixed points are unchanged.  
While this bifurcation does not affect the basin of attraction of the two stable fixed points, it has a large impact on the local field strength, as shown by the colored arrows.
The second pitchfork bifurcation takes place at $C=1/3$.
For this coupling value, each of the two new saddle fixed points splits into a stable fixed point and two further saddle points. 
Therefore, for very weak coupling we observe four basins of attractions, whose shape is shown in Fig. \ref{fig:pp1}.C.
The stability of the fixed points in $(w,z)=(-1,-1)$ and $(w,z)=(1,1)$ is not affected by the bifurcations (see appendix A).
On the other hand, if we increase the coupling coefficient, then the two nullclines will progressively flatten onto each other, and in the limit $C \rightarrow \infty$ the system becomes linear, with overlapping nullclines. 
These observations have been summarized in the bifurcation diagram of Fig. \ref{fig:bifurcations}.A. We observe that there are actually three pitchfork bifurcations, but that two of them are degenerate since they happen for the same value of $C$.

Note that, if the induction is null, $I=0$, changing the coupling $C$ or the potential strength $K$ has an equivalent effect on the nullclines, because only the ratio $K/C$ matters, cf. Eq. \ref{eq:nullclines}.
However, the field strength, defined as the norm of the vector $\left(\frac{dw}{dt}, \frac{dz}{dt}\right)$, is different.


\subsection{Asymmetric parameters choices can change the shape of the attraction basins}
\label{subsect:asymm_case}

\begin{figure}
\centering
\includegraphics[width=1.0\textwidth]{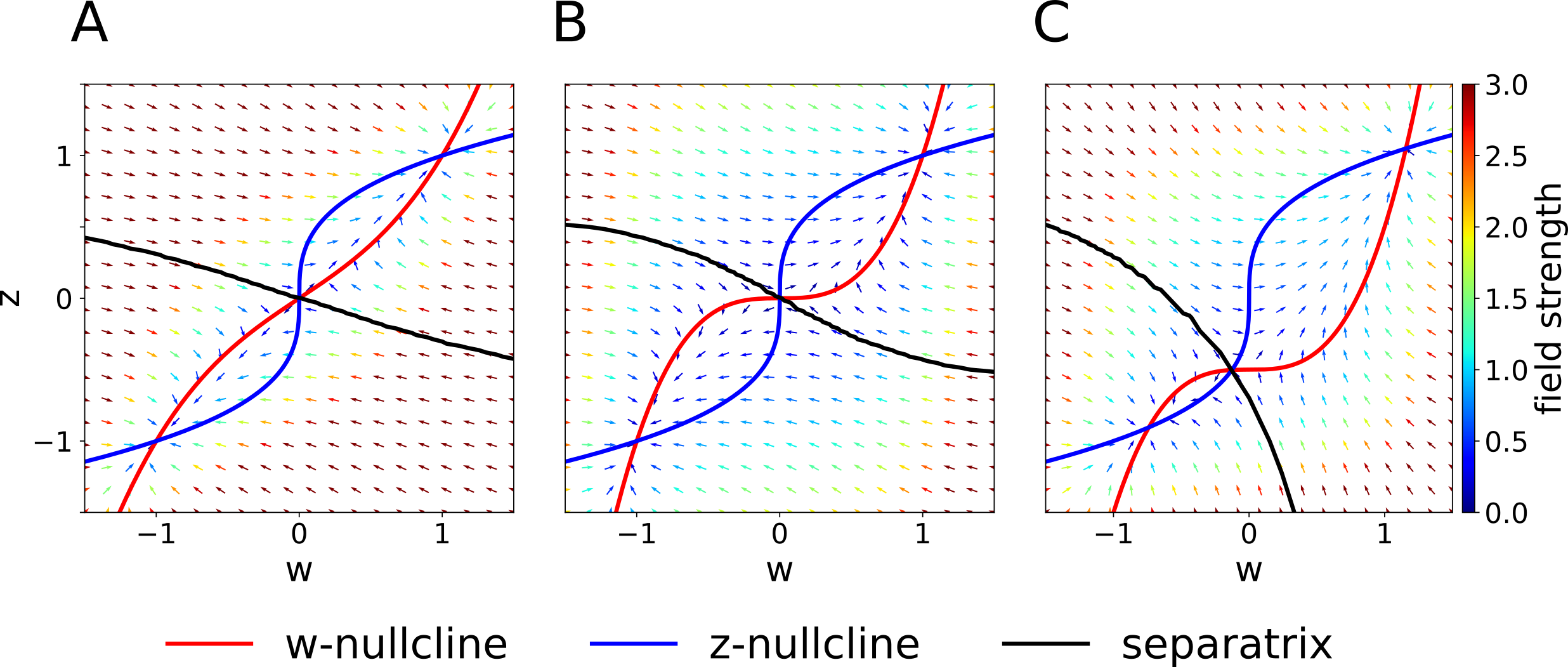}
\caption{\textbf{Asymmetric parameters choices.}  \textbf{A:} In the case $C_w=3>C_z=1$, the curvature of the $w-$nullcline is smaller than that of the $z-$nullcline and the basins of attraction are deformed compared to Fig. \ref{fig:pp1}.A. \textbf{B:} $\tau_z/\tau_w = 3$. Nullclines are not affected (compare to Fig. \ref{fig:pp1}.A) but the basin of attractions are. \textbf{C:} For $I=0.5$ (all other parameters set to 1), the basin of attraction of the fixed point at $(-1,1)$ is small compared to that of the fixed point at $(1,1)$.
}
\label{fig:asymmetry}
\end{figure}

As a more general case, we consider asymmetric $C$ or $\tau$.
When the coupling coefficients are asymmetric, we can plot the position of the bifurcation points in the $C_w$ - $C_z$ plane, as shown in Fig. \ref{fig:bifurcations}.B. The choice $C_w=C_z$ corresponds to the dashed gray line.
We notice that in the asymmetric case it is possible to have three distinct bifurcations. 
For example, we can fix $C_w=0.3$ and decrease $C_z$, from $1$ to $0$.
In the symmetric case, two bifurcations collapse into a single one.
The values of $C_w, C_z$ for which the first bifurcation happens can be easily computed analytically, as shown in Appendix A. In particular, by evaluating Eq. (\ref{bif_cond}) in the origin, we have the condition $C_w+C_z=1$. 
Therefore, if $C_w>1$, no bifurcation is possible and the number of fixed points is always three. 
Geometrically, changing the coupling coefficients $C_w$ or $C_z$  means changing the slope with which the nullclines intersect each other. This influences the shape of the attraction basins as shown in Fig. \ref{fig:asymmetry}.A.
On the other hand, if $C_w+C_z<1$, the system enters in the regime with minimum five fixed points.
Moreover, we can analytically compute the bifurcation value of one coupling constant, given the other.

If we keep $C_w=C_z$ but consider instead $\tau_z > \tau_w$, the system in Eq. (\ref{eq:dynamics}) may be interpreted as two different molecular mechanisms that act on different time scales. For example, the variable $z$ can be interpreted as a tagging mechanism or a consolidation variable.
A comparison of Fig. \ref{fig:pp1}  and Fig. \ref{fig:asymmetry}.B shows that the changes in $\tau$ do not affect the nullclines but change the field and the basin of attraction.



Another way by which we can introduce asymmetry in the system is by adding a plasticity induction current $I$. 
As it is clear from Eq. (\ref{eq:nullclines}), a positive induction $I> 0$ will cause a down shift of the $w-$nullcline. The symmetric case with $C_w=C_z=1$, $\tau_w=\tau_z=1$ s , $K_w=K_z=1$ and $I>0$ is shown in Fig. \ref{fig:asymmetry}.C.
The effect of $I>0$ also implies a reduction of the basin of attraction of the lower stable fixed point in favor of an increase of the attraction basin of the upper stable fixed point. 
For high values of induction, the attraction basin of the lower fixed point disappears leading to a bifurcation. 
Therefore, when $I>0$ is large enough,  the system is forced to move to the upper fixed point that can be interpreted as a potentiated state of the synapse.
Analogously, when $I<0$, the attraction basin of the lower fixed point is enlarged and leads, eventually, to a bifurcation in which the two upper fixed points are lost.

\vspace{1cm}
A possible generalization of the model would be to consider the coupling coefficients as dynamical variables. 
This possibility has been explored in other computational models \cite{ziegler2015}, with the use of gating variables. 
In this case, the coupling of the two dynamical variables  alternates, implementing a write-protection mechanism. 
The price we pay is the introduction of additional differential equations for the dynamics of the coupling coefficients and possibly more free parameters. 
In the specific implementation of \cite{ziegler2015}, the dynamical coupling represents a low-pass filter of the plasticity-inducing stimulus $I$ or the effect of neuromodulators on plasticity.

\section{Results}
\label{results}

In this section, we show that our two-dimensional model (see Method section for details) predicts a complex dependence of the synaptic consolidation dynamics upon the parameters of the experimental protocol.
This complex dependence has striking similarities with the behavior observed in experiments.    
First, we describe how we abstract the experimental protocol into a time-dependent input $I(t)$.
Then, we show the response of our model to different stimulation protocols.
After having analyzed the simpler case of a single rectangular pulse stimulation, we will move to the more realistic case of repetitive stimulation.
We will focus on synaptic potentiation, since synaptic depression is mirrored to the case of potentiation because of the symmetries of our model, since the self-interaction term in Eq. \ref{eq:dynamics}  is symmetric with respect to $w=z=0$.

\subsection{Abstraction of the stimulation protocol}
In their seminal work, Bliss and Lomo \cite{bliss1973long} showed that repeated, high-frequency stimulation of afferent fibers can lead to long-lasting synaptic potentiation.
On the other hand, in later work it was shown that low-frequency stimulation can lead to long-lasting synaptic depression \cite{bashir1994investigation}.
In order to keep the analysis transparent, we use a time-dependent, single-valued quantity $I(t)$ as an abstraction for the experimental protocol. 
In what follows, we will refer to $I(t)$ as the plasticity-induction current.
More precisely, we do not perform an explicit mapping from the electrical current used in LTP experiments for the stimulation of pre-synaptic fibers onto $I(t)$ that influences the dynamics of Eq. (\ref{eq:dynamics}), since this would require additional assumptions on how $I$ depends on the pre- and post-synaptic neural activity.
Instead, we model a set of high-frequency pulses as a single rectangular pulse of positive amplitude. The larger the stimulation frequency, the larger the amplitude. On the other hand, a set of current pulses at low frequency are modeled as a single negative rectangular pulse.
This choice is justified since the time between single pulses, even in the case of low-frequency stimulation, is very short compared to the timescale of plasticity.
This implies that multiple short pulses can be well approximated by a single longer pulse with the same area.
In agreement with well-established plasticity models \cite{pfister2006,clopath2010}, our choice of the sign of $I$ is dictated by the fact that high-frequency stimulation leads to potentiation as a positive $I$ leads to the loss of the depotentiated fixed point.
Conversely, a negative $I$ favors depotentiation.

\subsection{Stimulation using a single pulse}
\label{subsect:single_pulse}

\begin{figure}
    \centering
    \includegraphics[width=1\textwidth]{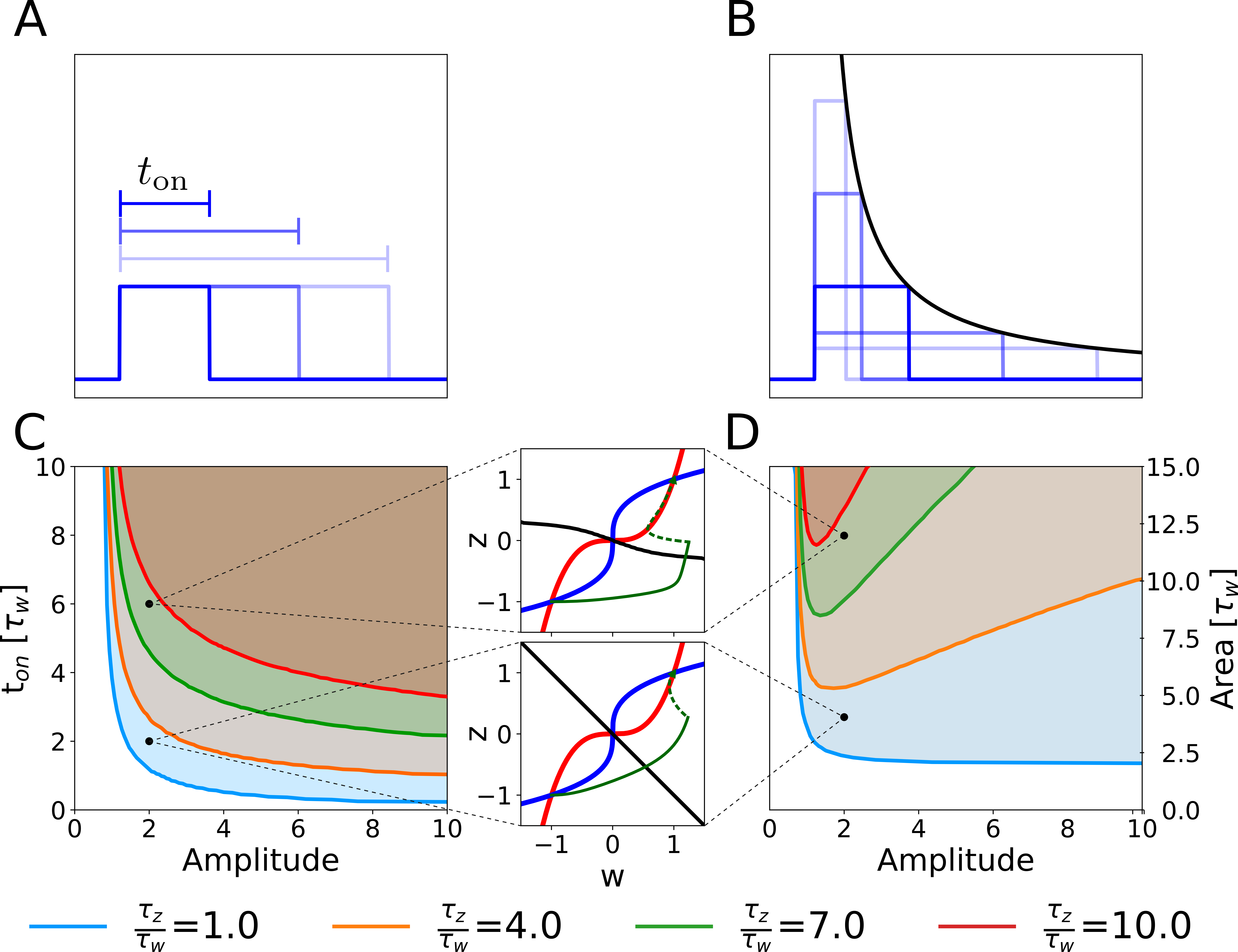}
    \caption{\textbf{Potentiation using single-pulse stimulation.}  Different curves correspond to different ratios of the time constant $\tau_z$ and $\tau_w$ in Eq. (\ref{eq:dynamics}). \textbf{A:} Schematic representation of single pulse stimulations, corresponding to different choices of $t_{\text{on}}$. \textbf{B:} Schematic representation of different single pulse stimulations with constant area. The black line is proportional to $1/amplitude$ in order to stress that all pulses have the same area. \textbf{C:} Separation curves between the potentiated and non-potentiated regions as a function of amplitudes and duration $t_{\text{on}}$ of a the single-pulse stimulus. The shaded region of the parameter space is the one in which the synapse gets potentiated. \textbf{D:} Same as \textbf{C}  but as a function of amplitude and area of the pulse. The two insets show examples of trajectories (green lines) in the phase-plane for two different parameters choices. The solid green lines represent the dynamical evolution of the system during the application of the external stimulus, while the dotted green line shows the relaxation of the system to a stable fixed point after the stimulation.}
    \label{fig:single_pulse}
\end{figure}

We consider the case in which our two-variable synapse model is stimulated with a single rectangular pulse of variable amplitude and duration $t_{\text{on}}$ (Fig. \ref{fig:single_pulse}A). 
Experimentally, this would correspond to compare high-frequency protocols with differing stimulation intensity (i.e. pulse frequency) and duration.
For each choice of duration and amplitude, we initialize the system in the $(w,z) = (-1,-1)$ state (depotentiated state) and we numerically integrate the system dynamics until convergence.
We then measure the final state of the synapse, i.e. whether it converged to the potentiated or the depotentiated state.
In Fig. \ref{fig:single_pulse} we plot the curve that separates the region of the parameter space that yields potentiation (shaded area) from the one that does not.
Different curves correspond to different time constants  $\tau_w$ and $\tau_z$ of the synaptic variables $w$ and $z$ in Eq. \ref{eq:dynamics}. 

Fig. \ref{fig:single_pulse}.C illustrate a rather intuitive result, i.e. if the amplitude of the pulse is increased, the duration needed for potentiation decreases. 
Moreover, if the amplitude is too small, we cannot achieve potentiation, even for an infinite pulse duration. 
The limit of infinite pulse duration is the ``DC'' limit, and the effect of this type of stimulation can be easily understood from the phase-plane analysis (Fig. \ref{fig:asymmetry}). 
Indeed, the introduction of a constant term in the $w$ equation (DC term), yields a shift in the $w$-nullcline vertically downward (if the term is positive). However, if the term is too small to cause the loss of the low fixed point, potentiation cannot be achieved (Fig. \ref{fig:single_pulse}.C).

The behavior of the separation curves in Fig. \ref{fig:single_pulse}.C might look like a power-law dependence, which would indicate that the relevant parameter for potentiation is the area under the rectangular pulse.
In the limit in which the amplitude goes to infinity and the duration goes to 0 while the area of the whole pulse stays the same, the stimulus can be described by a Dirac-$\delta$ function.
To study these aspects, we performed a similar analysis, this time varying the amplitude of the pulse and its area independently while fixing the duration to $\text{duration = \text{area}/\text{amplitude}}$ (see Fig. \ref{fig:single_pulse}.B).
The results are shown in Fig. \ref{fig:single_pulse}.D.
If there were a regime in which the relevant parameter is the area of the pulse, then the separation curve would be horizontal.
However, we find this type of behavior only for $\tau_z=\tau_w$ and in the high-amplitude region.
For  $\tau_z>\tau_w$ we find the existence of  an optimal value of the amplitude that yields potentiation with the minimal area. If we increase the amplitude beyond this optimal value, the necessary area under the stimulus curve $I(t)$ starts to increase again.
We can understand this effect by looking at the phase-plane, in particular at the dependence of the separatrix on the timescale separation.
In Fig. \ref{fig:asymmetry}.D we can see that, if $\tau_z \gg \tau_w$, the separatrix tends to an horizontal line for  $w\gg 1$. 
Since a $\delta$-pulse is equivalent to an instantaneous horizontal displacement of the momentary system state in the phase-plane, a single $\delta$-pulse cannot bring the system across the separatrix.
The $\delta$-pulse stimulation is, of course, a mathematical abstraction. In a real experimental protocol, such a stimulation can be approximated by a short stimulation pulse with short duration. Due to the finite duration of the short pulse, the system response in the phase-plane cannot be  a perfectly horizontal displacement. 
However, achieving potentiation with short pulses can still be considered as difficult, because it would require a disproportionate large stimulation amplitude.

Our findings highlight  the fact that different parameter sets yield very different behaviors of the model in response to changes in the stimulation protocols, which might be useful to design optimal experimental protocols. 
In particular, a model with timescale separation would predict the existence of an optimal pulse amplitude for which the total stimulus area necessary for potentiation is minimized.
We note that any model where consolidation works on a timescale that is slower than that of plasticity induction will exhibit timescale separation and be therefore sensitive to details of the stimulation protocol.

\subsection{Stimulation using repetitive pulses}
\label{rep_pulses}

\begin{figure}
\centering
\includegraphics[width=1.\textwidth]{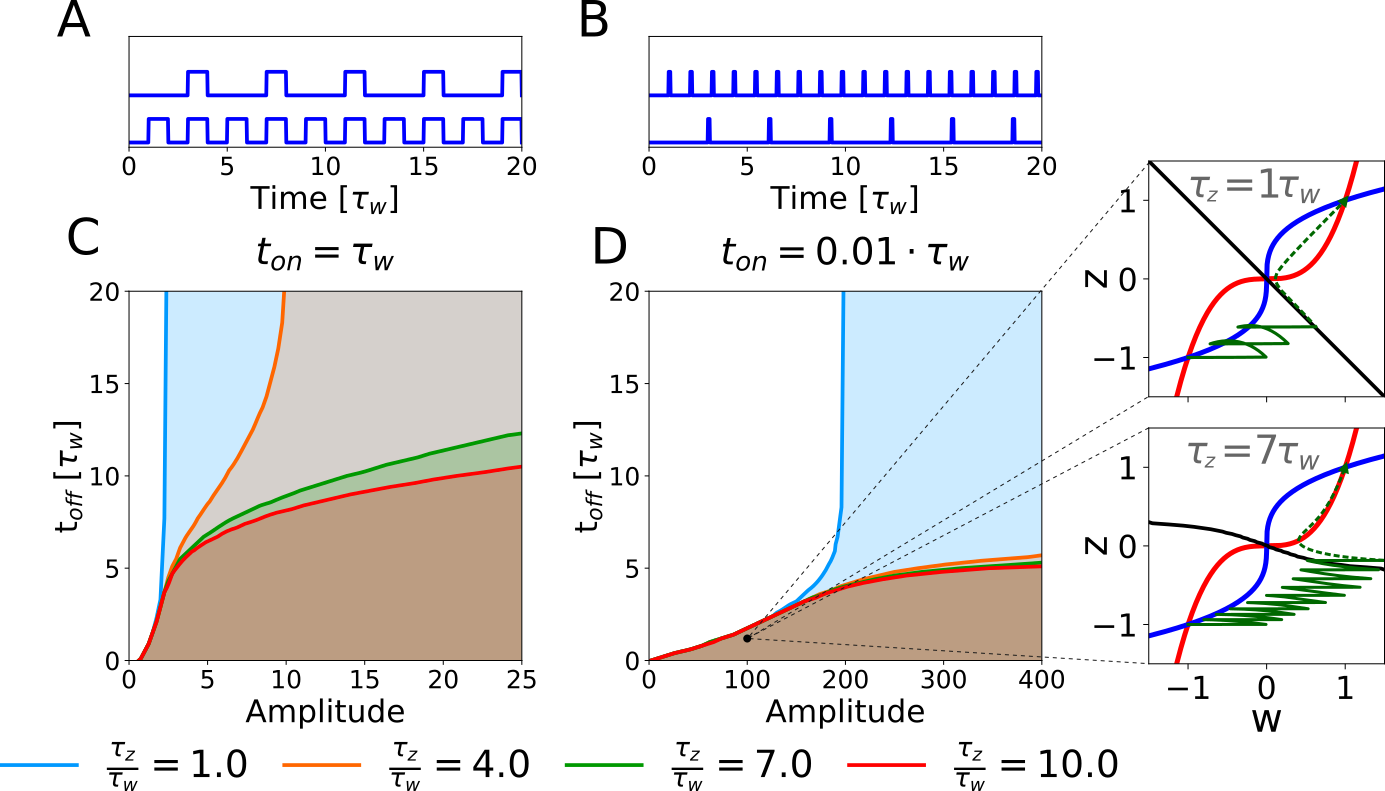}
\caption{Potentiation with repetitive stimulation. The curves indicate the separation between the region that yields potentiation (shaded) and the region that does not (white).
\textbf{A:} Schematic representation of stimulation protocols characterized by different $t_{\text{off}}$, while $t_{\text{on}}=\tau_w$ is fixed. 
\textbf{B:} Schematic representation of stimulation protocols characterized by different $t_{\text{off}}$, with $t_{\text{on}}=0.01\tau_w$. 
\textbf{C:} Potentiation region for long pulse stimulation, where $t_{\text{on}}=\tau_w$ is fixed.
\textbf{D:} Potentiation region for short pulse stimulation, with $t_{\text{on}}=0.01\tau_w$. The potentiation region is shaded. In comparison with \textbf{C}, the relevant amplitude interval changes dramatically depending on $t_{\text{on}}$.
The two insets show examples of trajectories (green lines) in the phase-plane for the same choice of stimulation parameters but different timescale separation. The solid green lines represent the dynamical evolution of the system during the application of the external stimulus, while the dotted green line shows the relaxation of the system to a stable fixed point after the stimulation.}
\label{fig:rep_pulses_fixed_ton}
\end{figure}

\begin{figure}
\centering
\includegraphics[width=1.\textwidth]{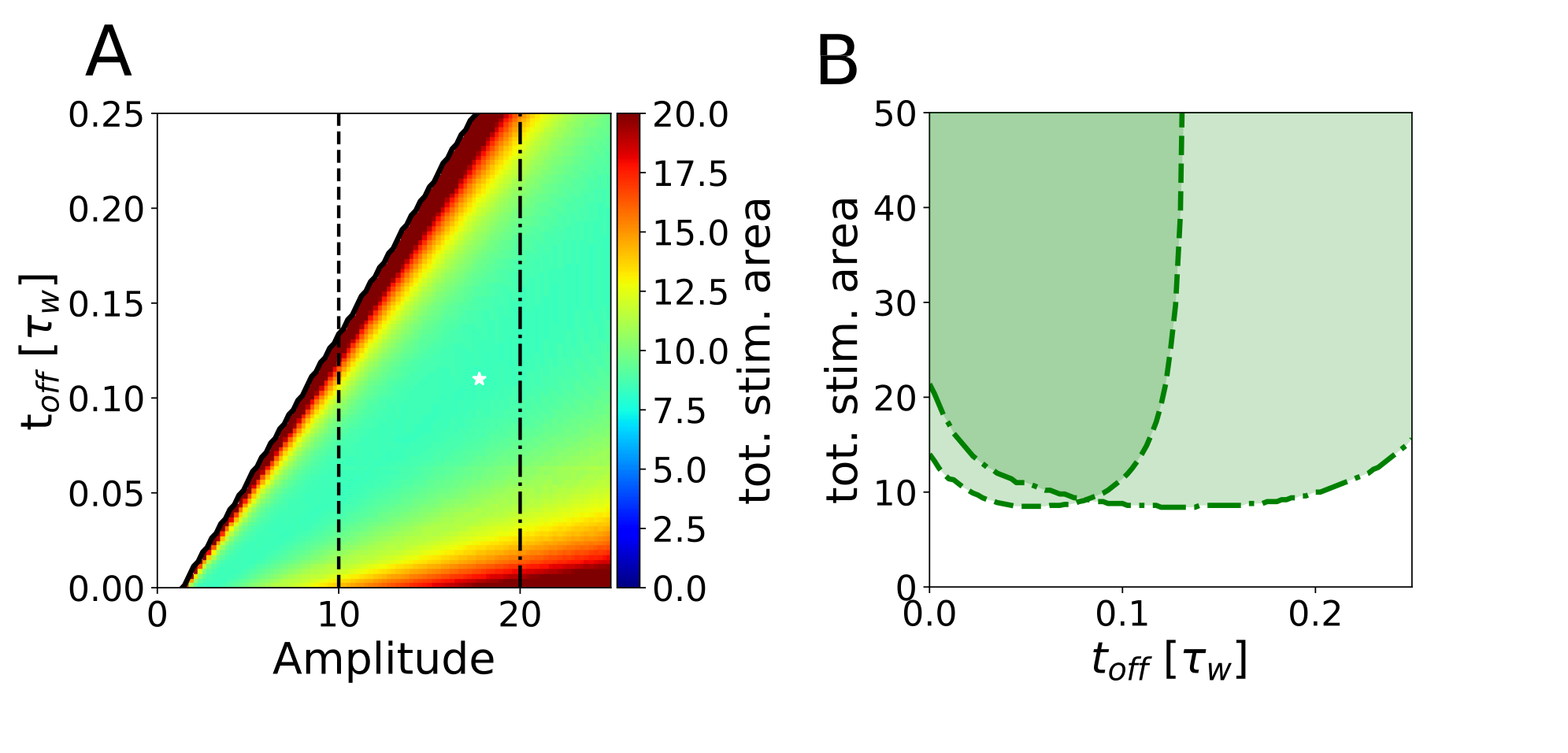}
\caption{Stimulation time need to achieve potentiation for $\tau_z/\tau_w=7$ and $t_{\text{on}}=0.01\tau_w$. \textbf{A.} In the potentiation domain (shaded in Fig. \ref{fig:rep_pulses_fixed_ton}) is colored in proportion to the stimulation area needed to  achieve potentiation with a repetitive pulse stimulus. The minimum stimulation area is $8.34$, it is indicated by the white star and corresponds to the parameters values $t_{\text{off}}=0.11 \tau_w$  and $amplitude=17.75$. \textbf{B.} The section of figure A for $amplitude =5$ (dotted line) is enlarged. One can notice that for a fixed stimulation amplitude, there is an optimal frequency value that minimizes the stimulus area required to achieve potentiation. Dotted and light-shaded: same for $amplitude=10$. }
\label{fig:rep_pulses_tot_area}
\end{figure}

As a second case, we consider the stimulation of a synapse by repetitive rectangular pulses. 
In an experimental setting, this type of stimulation would correspond to repetitive sets of high-frequency stimulations. 
This type of stimulus is characterized by three parameters: the amplitude of each pulse, its duration ($t_{\text{on}}$) and the time between pulses (or inter-pulse interval, $t_{\text{off}}$), as depicted in Fig. \ref{fig:rep_pulses_fixed_ton}.A,B.  
To keep the analysis transparent we consider a large enough number of repetitions to be sure whether potentiation is achieved or not given the three parameters.
Notice that if $t_{\text{off}}=0$ we are back to the constant stimulus case.

Fig. \ref{fig:rep_pulses_fixed_ton} shows the separation curves between potentiation (shaded) and no change (white) in the amplitude-$t_{\text{off}}$ space for fixed values of $t_{\text{on}}$ and for different $\tau_z/\tau_w$ ratios. 
In Fig. \ref{fig:rep_pulses_fixed_ton}.C we fix $t_{\text{on}} = \tau_w$.
We observe that, at least for low timescale ratios, it exists an amplitude above which the synapse gets potentiated independently of $t_{\text{off}}$, which suggests that potentiation happens at the first pulse. 
The amplitude necessary to obtain potentiation in one pulse, however, increases steeply with the $\tau_z/\tau_w$ ratio. 
On the other hand, if the value of $t_{\text{off}}$ is small enough (i.e. for high repetition frequency), potentiation can be achieved with smaller amplitudes and the timescale ratio is less important (notice the superimposed lines in the bottom left part of the plot).
If we decrease the pulse duration to $t_{\text{on}}=0.01\tau_w$ we obtain qualitatively similar separation curves, but potentiation now requires much larger values for the amplitude (see Fig. \ref{fig:rep_pulses_fixed_ton}.B), than for $t_{\text{on}}=\tau_w$ (see Fig. \ref{fig:rep_pulses_fixed_ton}.A).

In analogy to the analysis performed in section \ref{subsect:single_pulse}, we look for an optimal stimulation protocol in the case of repetitive pulses. 
In order to allow a direct comparison between single and repetitive pulse regimes, we measured the total area under the stimulation curve $I(t)$ in the repetitive pulse scenario limited to a finite number of pulses.
In Fig. \ref{fig:rep_pulses_tot_area}.A, we show the minimum stimulation area (number of pulses times the area of each pulse) required to achieve potentiation, as a function of the amplitude and the frequency of the stimulus for strong timescale separation ($\tau_z/\tau_w=7$). 
We notice that the minimum stimulation area (white star) corresponds to  $t_{\text{off}}\sim 10 \, t_{on}$. 
In real experimental conditions, however, it might be difficult to control the amplitude of the stimulation.
If we consider a fixed pulse amplitude, we find that there exists an optimal stimulation frequency to obtain potentiation with minimal total area (see Fig. \ref{fig:rep_pulses_tot_area}.B). 
This results highlight the fact that, for most stimulation amplitudes, one can find an optimal repetition frequency.

\section{Discussion}

We introduced and analyzed a minimal mathematical model of synaptic consolidation, that consists of two first-order ODEs with linear coupling terms and cubic nonlinearity.
Since it is a two-dimensional model, the system can be studied using  phase-plane techniques. 
While our model can have up to four stable fixed points, we focused on the case of two stable fixed points, to allow the physical interpretation of the fixed points as a depressed or potentiated synapse.

We showed that our minimal model responds to stimulation protocols in a non-trivial way, exhibiting a non-monotonic dependence of the necessary stimulation area on the stimulus parameters.
In particular, we found that for both single pulse and repetitive pulse stimulation it is possible to choose the stimulation parameter optimally, i.e. minimizing the stimulus total area.
We compared the minimum stimulation area needed to achieve potentiation with a single pulse to the case of repetitive pulses and we showed that, under most parameter values, the latter type of stimulation is advantageous to achieve potentiation.
Moreover, in experiments it is hard to have a fine control on stimulation amplitudes, because extra-cellular stimulation must be strong enough to be transmitted to the post-synaptic neuron and because there is no control on the post-synaptic firing, which could undergo adaptation or exhibit other time-dependent mechanisms.
For any value of the amplitude, our model predicts the existence of a finite optimal repetition frequency.
Analogously, we notice that for fixed repetition frequency, an optimal amplitude can be determined.
The minimum of the total power is particularly pronounced in the regime of strong separation of timescale, a regime which is in agreement with the experimental literature on synaptic consolidation which suggests multiple mechanisms with a broad range of time scales \cite{bliss1993}.
Assuming that the timescale $\tau_w$ is on the order of tens of seconds, as suggested by some plasticity induction experiments \cite{petersen1998all}, we can interpret a short stimulation pulse of duration $0.01\cdot \tau_w$ as a burst of few pulses at high frequency.
For example, one particularly interesting protocol is the theta burst simulation, consisting of 4 pulses at 100Hz every 200 ms \cite{larson2015theta}.
Assuming that this stimulation does not correspond to an extremely small amplitude value (a reasonable assumption since we are in the regime far away from the LTP threshold), our model predicts an optimal frequency (see Fig. \ref{fig:rep_pulses_tot_area}) on the order of $t_{\text{off}}= 0.11 \tau_w$, which is on the same order of the experimental result. 

The model can qualitatively reproduce the dynamics of the synaptic weights in slice experiments on synaptic consolidation \cite{malenka1991postsynaptic, bliss1993synaptic}.
However, since the model is for a single synapse, it cannot be applied to more complex experiments that involve cross-tagging, where effects of protein synthesis are shared between several synapses \cite{dudai2000, nader2000, redondo2011, frey1997}. 
Using our model we can make some qualitative predictions on experimentally measurable quantities. 
For example, by comparing the sub-figures C and D in Fig. \ref{fig:rep_pulses_fixed_ton}, we can see that the optimal stimulation parameters change by varying the single pulse duration $t_{\text{on}}$.
More precisely, our model would predict that for shorter $t_{\text{on}}$ the optimal stimulus requires a large pulse amplitude.

The framework proposed in this paper can be related to a number of previous modeling approaches to synaptic consolidation. 
The synaptic consolidation mechanism of the model presented in \cite{zenke2015} falls precisely into our framework. 
In particular, the choice of \cite{zenke2015} would correspond to fixing $K_w=0$ in Eq. (\ref{eq:dynamics}), which results in a linear differential equation for $w$.
We notice that in their model the coupling term also depends on the post-synaptic activity, effectively introducing a time-dependent coupling coefficient, a procedure similar to the gating variable discussed earlier.
The model presented in \cite{ziegler2015} belongs to the three-dimensional generalization of the framework we described in this paper.
The dynamical understanding of the interplay between stimulation protocol and autonomous dynamics gained here by studying the two-dimensional system can be also applied to a three-dimensional generalization (and even to higher dimensions), under the assumption that coupling exists only between pairs of variables and that there is timescale separation.  
Using such a multi-dimensional generalization, it would be possible to explain a much larger set of experimental results. 
However, the model presented in \cite{ziegler2015} also features coupling coefficients that are dynamically adjusted as a function of the induction protocol itself, as described in section \ref{subsect:asymm_case}, which makes it at the same time even more expressive and hard to analyze.

Another interesting comparison is with the cascade model \cite{fusi2005}.
To illustrate this comparison let us assume that our model has several slow variables $z_1,\dots, z_n$ with time constants $\tau_1,\dots, \tau_n$. The coupling from $k$ to $k+1$ is analogous to the coupling of $w$ to $z$ in Eq. (\ref{eq:dynamics}).
Even though this extended model and the cascade model share the concept of slower variables, there are some important differences between the two.
First, the model of \cite{fusi2005} is intrinsically stochastic, i.e. the stochasticity due to spiking events is combined with the stochasticity of plasticity itself.
Second, the transitions among states in \cite{fusi2005} are instantaneous. 
In our framework instead, even though there are discrete \textit{stable} states, the transitions need some time to happen and this is exactly why the frequency of repetitive stimulus matters in our model.

Similarly to \cite{fusi2005}, the  ``communicating vessels'' model proposed in \cite{benna2016} relies on multiple hidden variables. 
However, in contrast to \cite{fusi2005}, the  dynamics are in \cite{benna2016}  determined by continuous variables that obey continuous-time differential equations. 
If we truncate their model to a single hidden variable, the resulting dynamics fall into our framework, with the simple choice $C_w=C_z=0$.

\section*{Acknowledgments}

The authors would like to thank Tilo Schwalger for useful comments and discussions.
This research was supported by the Swiss national science foundation, grant agreement 200020\_165538.

\section{Appendix A: Analysis of the stability of the fixed points}
\label{appendix1}

In section (\ref{subsect:symm_case}) we show that, while decreasing the value of the coupling, we encounter two bifurcation values of the coupling constants. 
In this appendix, we present the mathematical constraints for a bifurcation. 

The bifurcations described in section (\ref{subsect:symm_case}) are local bifurcation, i.e. in which the stability of a fixed point is mutated as a consequence of a parameter change 
\cite{beer1995}.
Local bifurcations are mathematically characterized by a real part of the eigenvalues of the Jacobian matrix at a fixed point equal to zero. 
This indicates that the stability properties of the fixed point itself are mutating. 
There are  two different kinds of local bifurcations: in our case we have a steady state bifurcation, in which the eigenvalue is zero, in contrast to a  Hopf bifurcation, in which only the real part of the eigenvalue is zero. 

The Jacobian matrix of the dynamical system in Eqs.(\ref{eq:dynamics}) is given by
\begin{equation}
    J = 
         \begin{pmatrix}
              \frac{1}{\tau_w}\left( \frac{\partial F_w(w)}{\partial w}- \frac{z_0 C_w}{w_0} \right) & \frac{C_w}{\tau_w} \\
              \frac{C_z}{\tau_z}  &   \frac{1}{\tau_z}\left( \frac{\partial F_z(z)}{\partial z}-\frac{w_0 C_z}{z_0} \right) \\
         \end{pmatrix} \quad '
\end{equation}
where we  defined the functions $F_w(w)=-K_w(w-w_0)(w+w_0)w $ and $F_z=-K_z(z-z_0)(z+z_0)z $. 

Given a fixed point $(\bar{w},\bar{z})$, one can find the steady state bifurcations by imposing the condition that $\det(\textbf{J}) |_{(\bar{w},\bar{z})}=0$:
\begin{equation}
C_w C_z =
\left(\frac{\partial F_w(w)}{\partial w}\bigg|_{(\bar{w},\bar{z})}-\frac{z_0 C_w}{w_0}\right)
\left(\frac{\partial F_z(z)}{\partial z}\bigg|_{(\bar{w},\bar{z})}-\frac{w_0 C_z}{z_0}\right) \quad ,
\end{equation}
 which simplifies to
\begin{equation}
     \frac{z_0 C_w}{w_0} \cdot \frac{\partial g(z)}{\partial z}\bigg|_{(\bar{w},\bar{z})} + \frac{w_0 C_z}{z_0}\cdot \frac{\partial F_w(w)}{\partial w}\bigg|_{(\bar{w},\bar{z})}=\frac{\partial F_z(z)}{\partial z}\bigg|_{(\bar{w},\bar{z})}\frac{\partial F_w(w)}{\partial w}\bigg|_{(\bar{w},\bar{z})} \quad .
    \label{bif_cond}
\end{equation}

Notice that, the determinant is evaluated at a fixed point,  $(\bar{w},\bar{z})$ which is dependent of the choice of coupling parameters $C_w$ and $C_z$, see Eq. \ref{eq:dynamics}.
Therefore, given that there are two remaining free parameters $C_w$ and $C_z$, we have one degree of freedom.

For example, we can study the stability of the fixed point in $(0,0)$ given the extra condition that $C_w=C_z$, than the solution to equation (\ref{bif_cond}) is $C_w=1/2$.

\end{document}